\documentclass[
aps,
pra,
showpacs
amsmath,
amssymb,
preprint,
floatfix,
lengthcheck, 
superscriptaddress
]{revtex4-1}

\usepackage{graphicx}
\usepackage{mathtools}
\usepackage{braket}
\usepackage{hyperref}
\usepackage{bbold}
\usepackage{calrsfs}
\usepackage{dsfont}
\usepackage{cancel}
\usepackage{mathrsfs}
\usepackage{multirow}
\usepackage{latexsym}
\usepackage[normalem]{ulem}
\usepackage{float}

\usepackage{color}

\DeclareMathAlphabet{\pazocal}{OMS}{zplm}{m}{n}

\usepackage{bm}

\makeatletter \renewcommand\@make@capt@title[2]{%
\@ifx@empty\float@link{\@firstofone}{\expandafter\href\expandafter{\float@link}}%
\sffamily{\textbf{#1}}\@caption@fignum@sep#2 }
\usepackage{placeins} 

\begin{document}


\title{Transition from Lorentz to Fano Spectral\\ Line Shapes in Non-Relativistic Quantum Electrodynamics}

\author{Davis M. Welakuh}
\email[Electronic address:\;]{dwelakuh@seas.harvard.edu}
\affiliation{Harvard John A. Paulson School Of Engineering And Applied Sciences, Harvard University, Cambridge, Massachusetts 02138, USA}

\author{Prineha Narang}
\email[Electronic address:\;]{prineha@seas.harvard.edu}
\affiliation{Harvard John A. Paulson School Of Engineering And Applied Sciences, Harvard University, Cambridge, Massachusetts 02138, USA}


\begin{abstract}
Spectroscopic signatures associated with symmetric Lorentzian and asymmetric Fano line shapes are ubiquitous. 
Distinct features of Fano resonances in contrast with conventional symmetric resonances have found several applications in photonics such as optical switching, sensing, lasing, and nonlinear and slow-light devices. Therefore, it is important to have control over the generation of these resonances. In this work, we show through \emph{ab initio} simulations of coupled light-matter systems that Fano interference phenomena can be realized in a multimode photonic environment by strong coupling to the electromagnetic continuum. Specifically, we show that by effectively enhancing the light-matter coupling strength to the photon continuum in an experimentally feasible way, we can achieve a transition from Lorentzian to Fano lines shapes for both electronic and polaritonic excitations. 
An important outcome of switching between these spectral signatures is the possibility to control the Purcell enhancement of spontaneous emission alongside electromagnetically induced transparency which is a special case of Fano resonances. Switching from Fano back to a Lorentzian profile can be achieved by physically reducing the coupling strength to the continuum of modes. Our results hold potential for realizing tunable Fano resonances of molecules and materials interacting with the electromagnetic continuum within multimode photonic environments.
\end{abstract}

\maketitle


\section{Introduction}

In atomic, molecular and solid-state physics, the spectroscopic detection of radiation characterizes the response of the system by probing physical processes such as scattering, fluorescence, or absorption~\cite{flick2018a,ruggenthaler2017b}.  
The spectral line shapes that arise provide fundamental information about these physical processes. A commonly observed spectral line shape is the symmetric Lorentzian line shape that represents the finite radiative lifetimes of excited states. In other cases, asymmetric line shapes are usually observed such as Fano resonances in auto-ionization in atoms~\cite{ott2013}. Such asymmetric line shapes occur when a discrete quantum state interferes with a continuum band of states~\cite{fano1935,fano1961}. Fano resonances have been identified in different spectra (scattering, emission or absorption) of several systems such as metamaterials with weak asymmetric split ring resonators~\cite{lukyanchuk2010}, quantum dots and wires~\cite{franco2003}, plasmonic nanostructures~\cite{lukyanchuk2010,fan2010}, molecular systems~\cite{alhassid2006}, strong coupling between Mie and Bragg scattering in photonic crystals~\cite{rybin2009} and cavity electromagnonics~\cite{gollwitzer2021} and electrodynamics~\cite{geng2021}. Among the wide range of applications of Fano interferences, some of the relevant applications are chemical and biomedical sensing~\cite{hao2008}, lasing~\cite{yu2017}, switching~\cite{yu2014}, and directional scattering~\cite{sheikholeslami2011}.


Recent work has considered the transition from Lorentzian to Fano resonances~\cite{ott2013,tu2015,wang2016} in the context of weak light-matter coupling using a semi-classical description. Light-matter interaction in a confined electromagnetic environment can be controlled such that a regime is reached where the energy is coherently exchanged between light and matter. In the case where strong light-matter coupling is achieved, a photonic mode hybridizes with a matter resonance giving rise to novel eigenstates commonly known as polaritons which are separated in energy by the so-called Rabi splitting~\cite{skolnick1998}. These new states of matter have attracted interest in enhancing or modifying chemical landscapes~\cite{hutchison2012,feist2018,flick2020}, electrical conductivities~\cite{laussy2010,orgiu2015}, biological processes~\cite{coles2014}, polariton lasing~\cite{ramezani2007} and enhance charge and energy transport~\cite{zhong2016,feist2015}, among others. With such control of physical and chemical processes, strong light-matter coupling is a timely approach to investigate how Lorentzian line shapes can be tuned into Fano resonances~\cite{limonov2017}.

In this \emph{Article} we investigate a transition between two seemingly different spectroscopic signatures associated with symmetric Lorentzian and asymmetric Fano line shapes in strongly coupled light-matter systems. Our first-principles description of the coupled light-matter system is based on quantum electrodynamical density-functional theory (QEDFT)~\cite{ruggenthaler2017b,ruggenthaler2014,flick2018,flick2015} in the linear-response regime~\cite{flick2019}. Using this approach, we show distinctively two cases where, on the one hand, electronic excitations are tuned from Lorentzian into Fano line shapes and on the other hand, polaritonic resonances that arise due to strong light-matter coupling are also tuned into Fano line shapes. This seamless transition between spectroscopic signatures is achieved by strongly coupling the bare and hybridized matter system composed of discrete states to the continuum of states of the electromagnetic vacuum. In order to confirm that the asymmetric line shapes that arise in the spectrum are indeed Fano resonances, we employ the Fano formalism~\cite{limonov2017,ott2013} that characterizes whether a profile is an asymmetric Fano or symmetric Lorentz line shape. We obtain excellent agreement between the fitted and \emph{ab initio} results which confirms that the asymmetric line shapes that arise are indeed Fano resonances. 

Next, we identify cases where the perturbative Fano formalism is inadequate to describe the coupled system, thus strengthening the need for an \emph{ab initio} description of strongly coupled light-matter systems. Furthermore, a particularly important physical outcome of strong coupling to the photon continuum is the Purcell enhancement of spontaneous emission where the finite radiative lifetimes of excited states get modified which is useful for realizing optical quantum memory devices~\cite{lvovsky2009}. The accompanying flexibility in tuning Fano resonances provides an approach to create sensing devices with variable characteristics, valuable for applications in photonics~\cite{limonov2021,limonov2017}. We also show the special case of Fano resonances where electromagnetically induced transparency is realized. The reverse process of switching from a Fano to a Lorentzian profile can be achieved by reducing the coupling strength to the continuum of modes. Our findings provide a valuable way to tune between the two different line shapes and would be applicable to scenarios where strong light-matter coupling can be enhanced within multimode photonic environments.

\section{Theoretical framework}
\label{sec:general-framework}

For the quantum mechanical description of light-matter interaction, we consider the Pauli-Fierz Hamiltonian in the long-wavelength limit~\cite{tannoudji1989,craig1998}. This implies that for a coupled matter-photon system, 
the relevant photon modes have wavelengths that are large compared to the size of a matter subsystem, for instance, an atom or molecule. In this setting, the length form of the Pauli-Fierz Hamiltonian in dipole approximation~\cite{rokaj2017,flick2019,rivera2019,ashida2021} is given by
\begin{align} 
\hat{H} &=\sum\limits_{i=1}^{N}\left(\frac{\hat{\textbf{p}}_{i}^{2}}{2m} + v_{\textrm{ext}}(\hat{\textbf{r}}_i)\right) + \frac{e^2}{4 \pi\epsilon_0}\sum\limits_{i>j}^{N}\frac{1}{\left|\hat{\textbf{r}}_i-\hat{\textbf{r}}_j\right|}\nonumber\\
& \quad +\sum_{\alpha=1}^{M}\frac{1}{2}\left[\hat{p}^2_{\alpha}+\omega^2_{\alpha}\left(\hat{q}_{\alpha} \!-\! \frac{\boldsymbol{\lambda}_{\alpha}}{\omega_{\alpha}} \cdot \hat{\textbf{R}} \right)^2\right].\label{eq:el-pt-hamiltonian}
\end{align}
Here, the $N$ electrons are described by the electronic coordinates, $\hat{\textbf{r}}_{i}$, and their corresponding momentum by the operator $\hat{\textbf{p}}_{j}$. The external potential of the nuclei is represented by $v_{\textrm{ext}}(\hat{\textbf{r}}_i)$ and $\hat{\textbf{R}}=\sum_{i=1}^{N}e \, \hat{\textbf{r}}_i$ represents the total electronic dipole operator where $e$ is the electron charge. The quantized electromagnetic field is described by harmonic oscillators which consists of the displacement coordinate $\hat{q}_{\alpha}$ and canonical momentum operator $\hat{p}_{\alpha}$ with associated mode frequency $\omega_{\alpha}$ for each mode $\alpha$ of an arbitrarily large but finite number of photon modes $M$. The physical coupling between light and matter is $\boldsymbol{\lambda}_{\alpha} = \frac{1}{\sqrt{\epsilon_{0}}} \, S_{\alpha}(\textbf{r}_{0}) \, \textbf{e}_{\alpha}$ where $\textbf{e}_{\alpha}$ is the transversal polarization vector of a photonic mode $\alpha$ with wave vector $\textbf{k}_{\alpha}$ and $\epsilon_{0}$ is the vacuum permittivity. The coupling strength depends on the form of the mode functions $S_{\alpha}(\textbf{r}_{0})$ and the chosen reference point for our matter subsystem~\cite{milonni1993}. Since we simulate the continuum of states by explicitly treating many photon modes, it might become necessary to use the \textit{bare mass} $m$ instead of the already renormalized physical mass $m_{e}$ (bare plus the electromagnetic masses) of the charged particles~\cite{spohn2004,rokaj2020}. However, we make the common assumption to treat explicitly the small part of the photon continuum that is changed with respect to the free space case due to the multimode photonic environment. This implies that the rest of the continuum of modes not affected by the photonic cavity is subsumed in the usual renormalized physical
mass (i.e. $m_{e}=1$ in atomic units) of the charged particles~\cite{spohn2004}. 

\begin{figure}[t] 
\centerline{\includegraphics[width=0.4\textwidth]{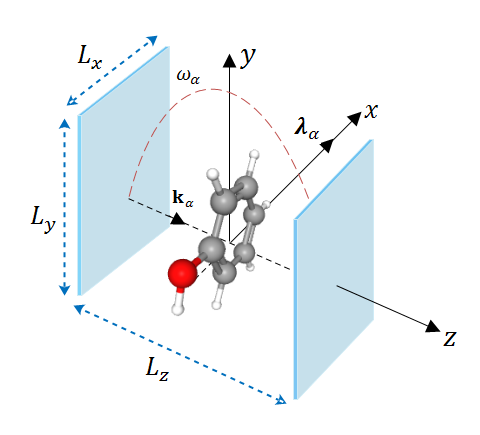}}
\caption{Schematic setup of a phenol molecule confined within a multimode optical cavity of volume $V=L_{x}L_{y}L_{z}$. The photon field is polarized along the $x$-axis with coupling strength $\boldsymbol{\lambda}_{\alpha}$ and the wave vector is along the cavity axis of length $L_{z}$ in the $z$-direction. The multimode cavity has a range of frequencies $\omega_{\alpha}$.}
\label{fig:phenol-in-cavity}
\end{figure}

We schematically depict in Fig.~(\ref{fig:phenol-in-cavity}) a physical setup to realize strong light-matter coupling featuring a single phenol (C$_{6}$H$_{5}$OH) molecule confined within a multimode optical cavity. Although we depict a standard Fabry-P\'{e}rot configuration, several other multimode photonic setups for strong light-matter coupling such as hybrid cavity-antenna configurations~\cite{zhang2019}, nanoparticle arrays~\cite{wang2018}, or coupled photonic crystals~\cite{dousse2010} are suitable for this study. Thus, the results presented here are not restricted to any specific photonic environment and not limited to a specific matter system (as demonstrated in App.~\ref{app:EIT-section}). To describe the coupling of the phenol molecule to the electromagnetic continuum, we sample a wide range of photon modes of a quasi one-dimensional mode space by employing the coupling strength $\boldsymbol\lambda_\alpha = \sqrt{\frac{2}{\epsilon_0 L_xL_y L_z}} \text{sin}(k_\alpha  \, z_0)\textbf{e}_x$, where $z_0 = L_z/2$ is the position of the molecule inside the cavity, $k_{\alpha}=\omega_{\alpha}/c$ and $\omega_\alpha = \alpha c\pi /L_z$ are the wave vectors and frequencies of the modes, respectively. We assume a constant mode function in the $x$- and $y$-direction and choose the volume $V=L_{x}L_{y}L_{z}$ (with $L_{z}=3250$ $\mu$m, $L_{x}=10.58$ \AA ~and $L_{y} =2.65$ \AA) such that the sampled modes couple weakly to the molecular system~\cite{flick2019,flick2017}. The electromagnetic vacuum is sampled with 100,000 photon modes with energies that cover densely a range from 0.19~meV for the smallest energy up to 38.14~eV for the energy cutoff with an equal spacing of $\Delta\omega= 0.38$~meV. Since we sample frequencies corresponding to a one-dimensional mode space, the lifetimes will differ from the actual three-dimensional lifetimes, but for the purpose of demonstrating the
possibilities of tuning between spectroscopic signatures, this is suﬃcient. In this setting, the coupling strength $\boldsymbol{\lambda}_{\alpha}$ can be increased by reducing the volume $V$ of the cavity which can be realized experimentally. Since we sample a one-dimensional mode space with constant mode functions in the $x$- and $y$-direction, we choose to change the coupling strength by varying the area $L_{x}L_{y}$.

As mentioned previously, when discrete quantum states interfere with a continuum of states this give rise to Fano line shapes that manifest, for example, in the absorption spectrum. This was realized in the original work of Fano when atomic levels were coupled to the continuum~\cite{fano1935,fano1961}. In our setting, the discrete states can be electronic or polaritonic that couple to the electromagnetic continuum, which makes it a viable setup to obtain Fano resonances for a large variety of systems. One of the methods to identify and characterize a Fano resonance in any spectra is to use the Fano formula~\cite{fano1935,ott2013}, which for the photo-absorption cross-section at energy $E = \hbar\omega$ is given as follows
\begin{align}
\sigma(E) = \sigma_{0} \frac{(q + \varepsilon)^{2}}{1 + \varepsilon^{2}} \, , \label{eq:cross-section-Fano-formula}
\end{align}
where $\sigma_{0}$ is the resonant cross-section, the dimensionless energy is $\varepsilon=2(E - E_{0})/\hbar\Gamma$, with $E_{0}$ being the resonant Fano energy and $\Gamma$ is the spectral line width. The parameter $q$ is the so-called Fano parameter that describes the degree of asymmetry of a Fano line shape. This parameter depends on the phase angle $\varphi$ as $q=\cot \varphi$ (and is periodic in $\pi$). The angle $\varphi$ describes the phase shift of the continuum. In the limit $q \rightarrow \pm \infty$ when $\varphi \rightarrow n\pi$ ($n$ is an integer) leads to Lorentzian line shapes and Fano resonances are obtained between these extreme cases where $q$ is finite and non-zero~\cite{limonov2017,limonov2021}. When $q=0$, the Fano profile is a symmetric quasi-Lorentzian antiresonance which physically represents the case when an external perturbation does not couple to the discrete state. The phase shift $\varphi$ can be understood as an interference of two transition amplitudes from initial to final state. In this context, the dimensionless Fano parameter $q$ is the ratio of the amplitude representing a transition to the discrete state and the amplitude describing the transition to the continuum~\cite{miroshnichenko2010}. The Fano formula (\ref{eq:cross-section-Fano-formula}) is generally applicable not only to absorption but also to different optical spectra (such as transmission and scattering) in a variety of systems~\cite{limonov2017}. Here we study the spectra by considering the dipole strength function which is related to the photo-absorption cross-section as $S(E)=(m_{e}c/2\pi^{2}e^{2}\hbar^{2}) \, \sigma(E)$~\cite{yabana1999pra}. In addition to Fano resonances that arise due to discrete to continuum coupling, two other distinct interference effects occur: (i) a change of spontaneous emission caused by the Purcell effect and (ii) a shift in energy levels associated with the Lamb shift. These interference effects will be discussed next.

\section{Results and Discussion}
\label{sec:results-and-discussion}

In this section, we investigate how to tune Lorentzian resonances into Fano line shapes in a cavity QED setup of a molecule coupled to the electromagnetic continuum. We study this physical effect in the absorption spectrum of the coupled system where Lorentzian line shapes of electronic and polaritonic excitations can be tuned. To study the coupled electron-photon system, we solve the stationary eigenvalue problem of Eq.~(\ref{eq:el-pt-hamiltonian}) by employing the reformulation of linear-response theory for light-matter systems within the framework of QEDFT~\cite{flick2019}. In particular, we use the electron-photon Casida approach introduced in Ref.~\cite{flick2019} while noting that similar approaches as in Refs.~\cite{flick2018,welakuh2022} can be used. For the spectrum considered here, the electronic and polaritonic excitations (discrete states) are within the energy cutoff of the photon continuum (large photonic vacuum density of states) which is necessary to generate Fano resonances.

\begin{figure}[bth]
\centerline{\includegraphics[width=0.5\textwidth]{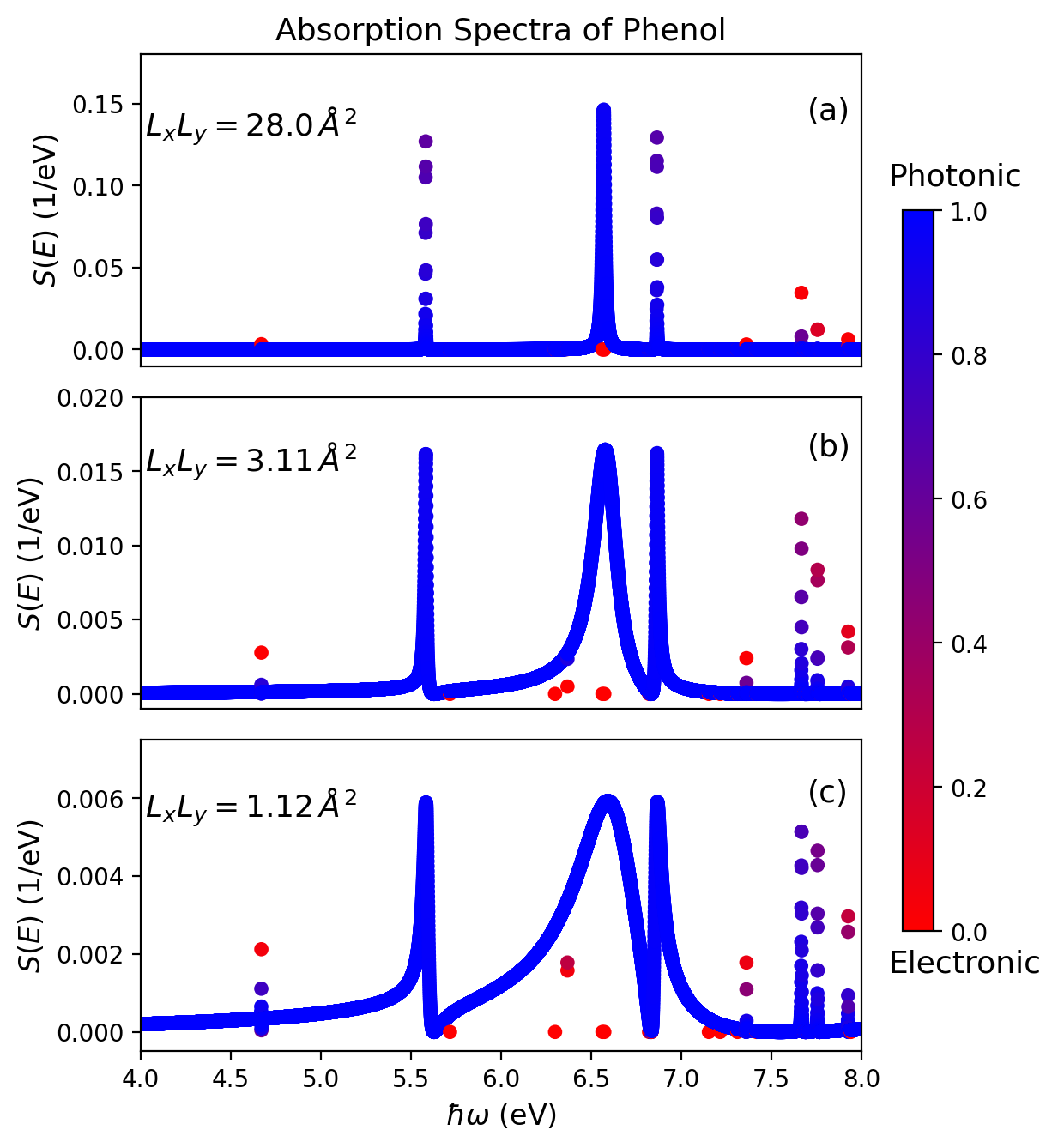}}
\caption{(a) \emph{Ab initio} calculation of the electronic excitation spectrum of a phenol molecule weakly coupled to cavity modes of a multimode photonic environment. Panels (b) and (c) show how the Lorentzian resonances are tuned into Fano resonances by coupling strongly to the electromagnetic continuum. }
\label{fig:phenol-electronic-spectrum}
\end{figure}

In the first numerical example, we couple the phenol molecule weakly to the photon continuum (as discussed in Sec.~\ref{sec:general-framework}). The results of this calculation are shown in Fig.~(\ref{fig:phenol-electronic-spectrum}). We note that the peaks are not artificially broadened in this calculation since we have sampled densely the photon continuum that couples to the matter subsystem. This case specifies an \emph{ab initio} approach on how to obtain lifetimes non-perturbatively from QEDFT linear-response theory as discussed in Ref.~\cite{flick2019}. The color code in Fig.~(\ref{fig:phenol-electronic-spectrum}) indicates the different contributions of each excitation in the response function where from blue means more photonic to red more electronic. Since the peak strengths are dominantly blue indicates that resonances are mainly photonic in nature and the peak positions are mostly dominated by the matter constituents. This result is in agreement with measurements of an absorption (or emission) spectrum where a photon is absorbed (or emitted) and not the matter constituents. In Fig.~(\ref{fig:phenol-electronic-spectrum} a), we show the spectrum of the molecule where we find Lorentzian line shapes for the different excitations. The light-matter coupling strength in Fig.~(\ref{fig:phenol-electronic-spectrum} a) corresponds to weak coupling to the photon continuum (for $L_{x}L_{y}=28.0$ \AA$^{2}$). When the light-matter coupling strength is effectively increased by reducing the volume of the cavity along the $x$- and $y$-direction, the molecule couples strongly to the photon continuum and this changes the spectrum. For reducing $L_{x}L_{y}$ (and increasing $\boldsymbol{\lambda}_{\alpha}$) as in Fig.~(\ref{fig:phenol-electronic-spectrum} b,c), we find a transition of the initially Lorentzian line shapes for the weak coupling to the continuum to Fano line shapes for strong coupling to the photon continuum. The asymmetric line shapes are due to the overlap of the discrete states of the electronic system with the continuum states of the electromagnetic vacuum, where destructive and constructive interference take place at close energy positions. 

\begin{table}[bth]
\begin{tabular}{ | c | c | c | c | }
	\hline
	$\; L_{x}L_{y}$ (\AA$^{2}$) & $\; \varphi$ (rad) &  $\; \hbar\Gamma$ (eV)  & $ \; q$  \\
	\hline 
	28.0  &     $\pi$    &   0.003   &  -8.17$\times 10^{15}$      \\
	\hline
	3.11  &  $0.972\pi$  &   0.015   &   -11.43    \\
	\hline
	1.12  &  $0.906\pi$  &   0.041    &   -3.27     \\
	\hline
\end{tabular}
\caption{Fitted values for the electronic case of the phase shift $\varphi$, line width $\hbar\Gamma$ and Fano parameter $q$ for the different cases of changing the coupling strength to the continuum by varying the cavity mode volume via $L_{x}L_{y}$. This corresponds to the Fano fit in Fig.~\ref{fig:phenol-electronic-spectrum-fit}.}
\label{tab:electronic-fit}
\end{table}
%

\begin{figure}[bth]
\centerline{\includegraphics[width=0.5\textwidth]{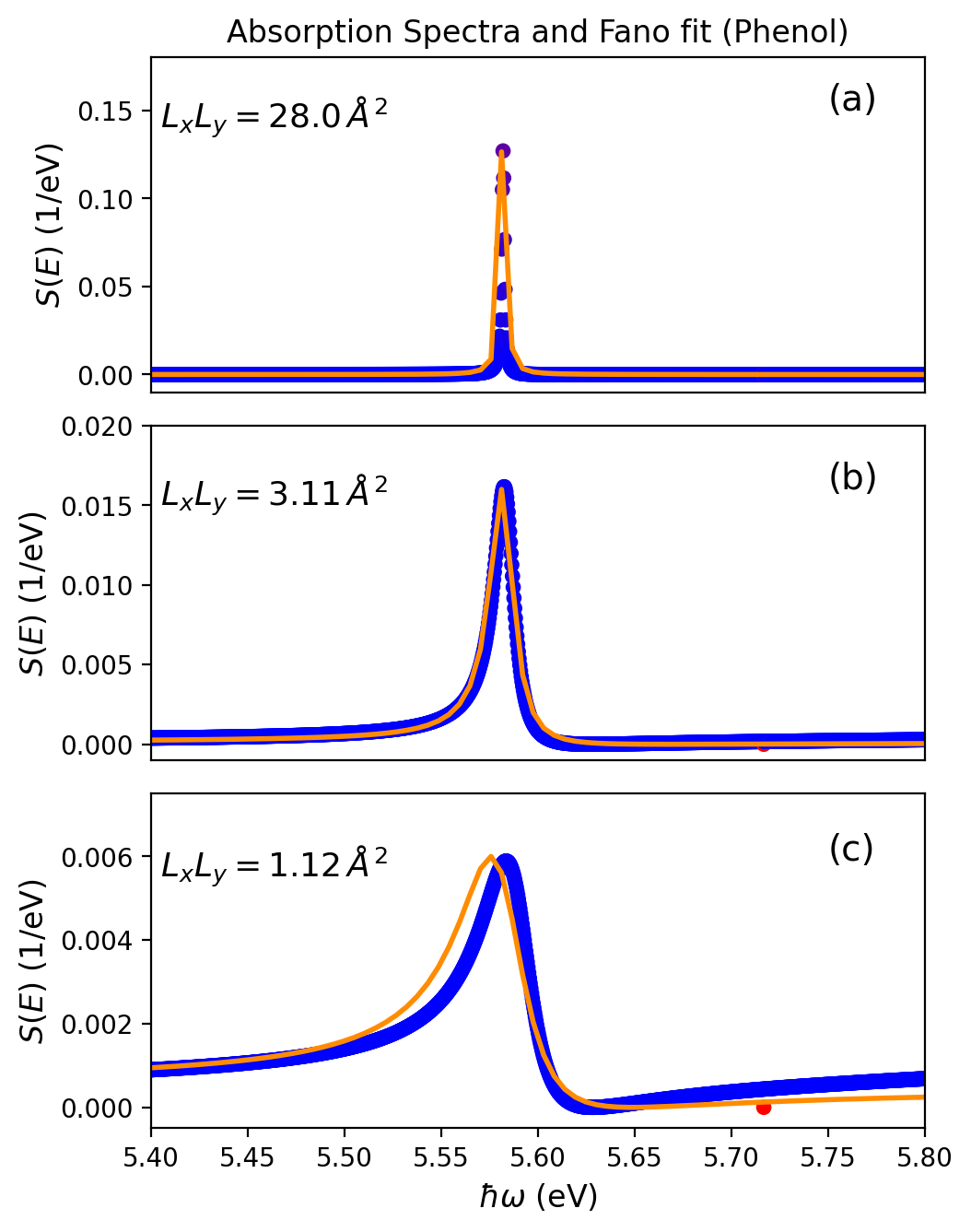}}
\caption{A Fano fit (orange line) to the peak around $5.6$~eV of Fig.~(\ref{fig:phenol-electronic-spectrum}) characterizing a transition from a symmetric to an asymmetric line shape. (a) Symmetric Lorentzian line shape occur in the weak coupling to the photon continuum. In panel (b) and (c), the asymmetric line shapes emerge due to strong coupling to the photon continuum as characterized by the Fano fit.}
\label{fig:phenol-electronic-spectrum-fit}
\end{figure}

To identify the physics of Fano resonances in the calculated spectra, we fit the \emph{ab initio} results with the Fano formula of Eq.~(\ref{eq:cross-section-Fano-formula}) by varying different parameters (see Tab.~\ref{tab:electronic-fit} for fitted values). We do this specifically for the peak around $5.6$~eV as shown in Fig.~(\ref{fig:phenol-electronic-spectrum-fit}).
In Fig.~(\ref{fig:phenol-electronic-spectrum-fit} a), we find a symmetric Lorentzian line shape for the weak coupling to the photon continuum as characterized with the Fano fit where the $\pi$ phase shift result to an infinite Fano parameter $q$. The symmetric Lorentzian shape is due to the negligible transition amplitude for weak coupling to the continuum. Fitting to the Lorentzian line shape is consistent with the Wigner-Weisskopf theory~\cite{weisskopf1930,flick2019} as the weakly interacting continuum modes merely describes the emission profile. For the strong coupling case, the Fano fit agrees with the \emph{ab initio} results in Fig.~(\ref{fig:phenol-electronic-spectrum-fit} b) and deviates slightly in  Fig.~(\ref{fig:phenol-electronic-spectrum-fit} c). The asymmetric line shape arises since the transition amplitude to the continuum increases for stronger coupling such that both amplitudes (transition to discrete and continuum) are different from zero, thus, giving a finite $q$. The competing interference effect that characterizes the asymmetric Fano resonance manifests itself as an increase in the cross section on one of the slopes of the profile (constructive interference) and a decrease on the second slope (destructive interference). This interference phenomena also explains the reduced oscillator strength for increasing coupling to the continuum. As changes in the electronic and photonic subsystem are self-consistent, the deviation of the Fano fit in Fig.~(\ref{fig:phenol-electronic-spectrum-fit} c) will become more pronounced for stronger coupling situations. This is expected since Fano used a perturbation approach to explain the appearance of asymmetric resonances~\cite{fano1961,miroshnichenko2010}. Typically, since the photonic density of states now differ from the vacuum density of states due to an increase in the coupling strength, this usually result in the enhancement (or suppression) of the rate of spontaneous emission~\cite{burstein1995}. In our example, the line width becomes broader (see Fig.~(\ref{fig:phenol-electronic-spectrum})), implying a shortened lifetime which can be obtained from Tab.~(\ref{tab:electronic-fit}) using $\tau=1/\Gamma$. The competing constructive and destructive interference that leads to a decrease of the oscillator strength gives rise to an enhancement of the radiative decay rate and a broadened line width~\cite{rybin2016}. This has potential applications for realizing a optical quantum memory device~\cite{lvovsky2009}. Another interesting effect that arise is the Lamb shift due to the interaction of discrete electronic states with the vacuum fluctuations of electromagnetic field. It is not ideal to associate the shift of the calculated spectra from the Fano fit in Fig.~(\ref{fig:phenol-electronic-spectrum} b,c) since for one, the Fano formalism is perturbative. However, the \emph{ab initio} results presented in this work are non-perturbative and we note that the shift can be partly attributed to the Lamb shift as discussed in Refs.~\cite{flick2019,welakuh2021t} and also to the photon random-phase approximation applied to the transverse electron-photon interaction terms.

\begin{figure}[bth!]
\centerline{\includegraphics[width=0.5\textwidth]{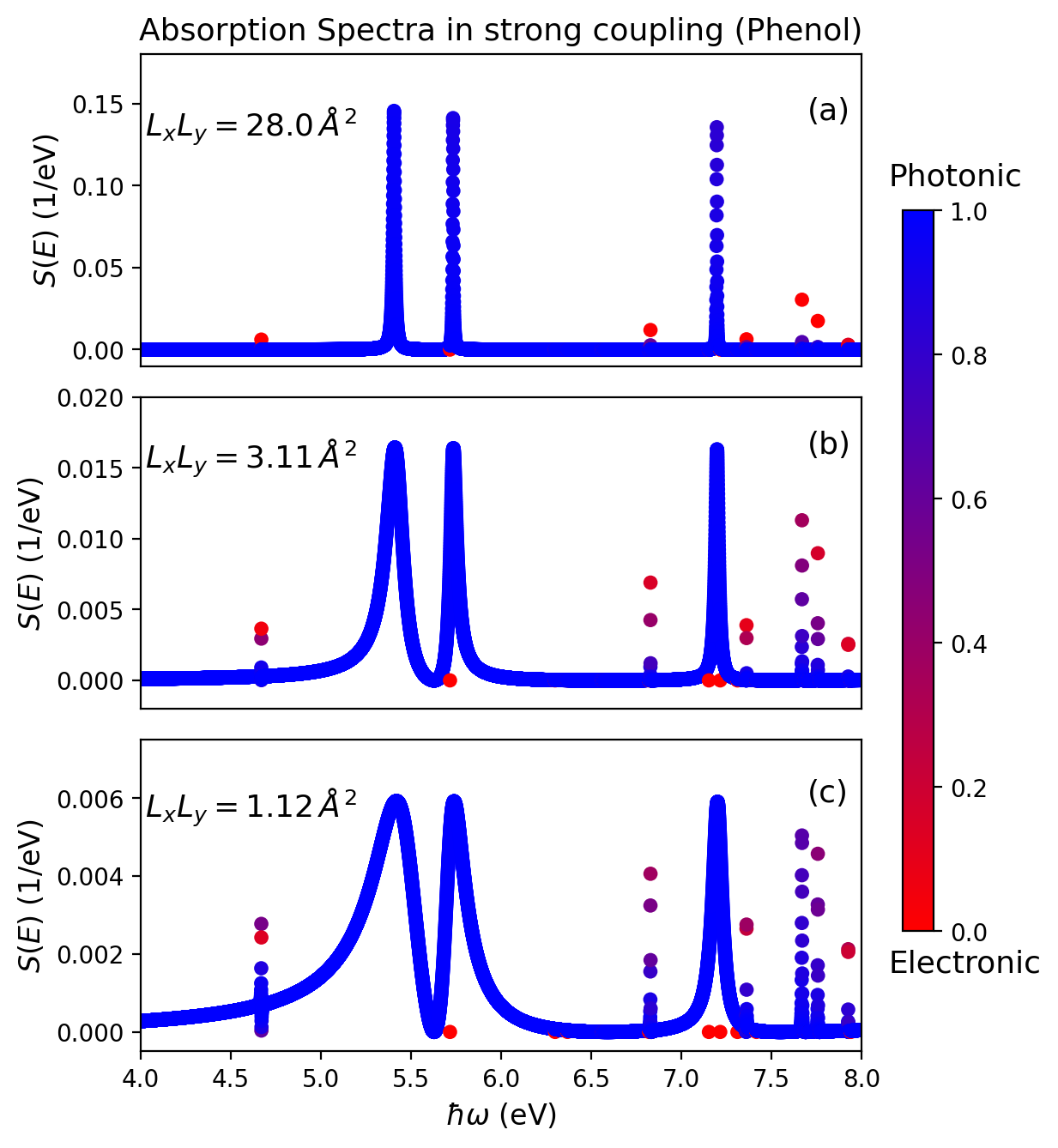}}
\caption{(a) \emph{Ab initio} calculation of the absorption spectrum of phenol capturing the lower and upper polaritonic peaks between 5 and 6~eV due to resonant strong coupling where the peaks are Lorentzian. Panels (b) and (c) shows how both polaritonic and electronic excitations can be tuned into Fano resonances by strongly coupling to the photon continuum.}
\label{fig:phenol-polaritonic-spectrum}
\end{figure}

\begin{figure}[ht] 
\centerline{\includegraphics[width=0.5\textwidth]{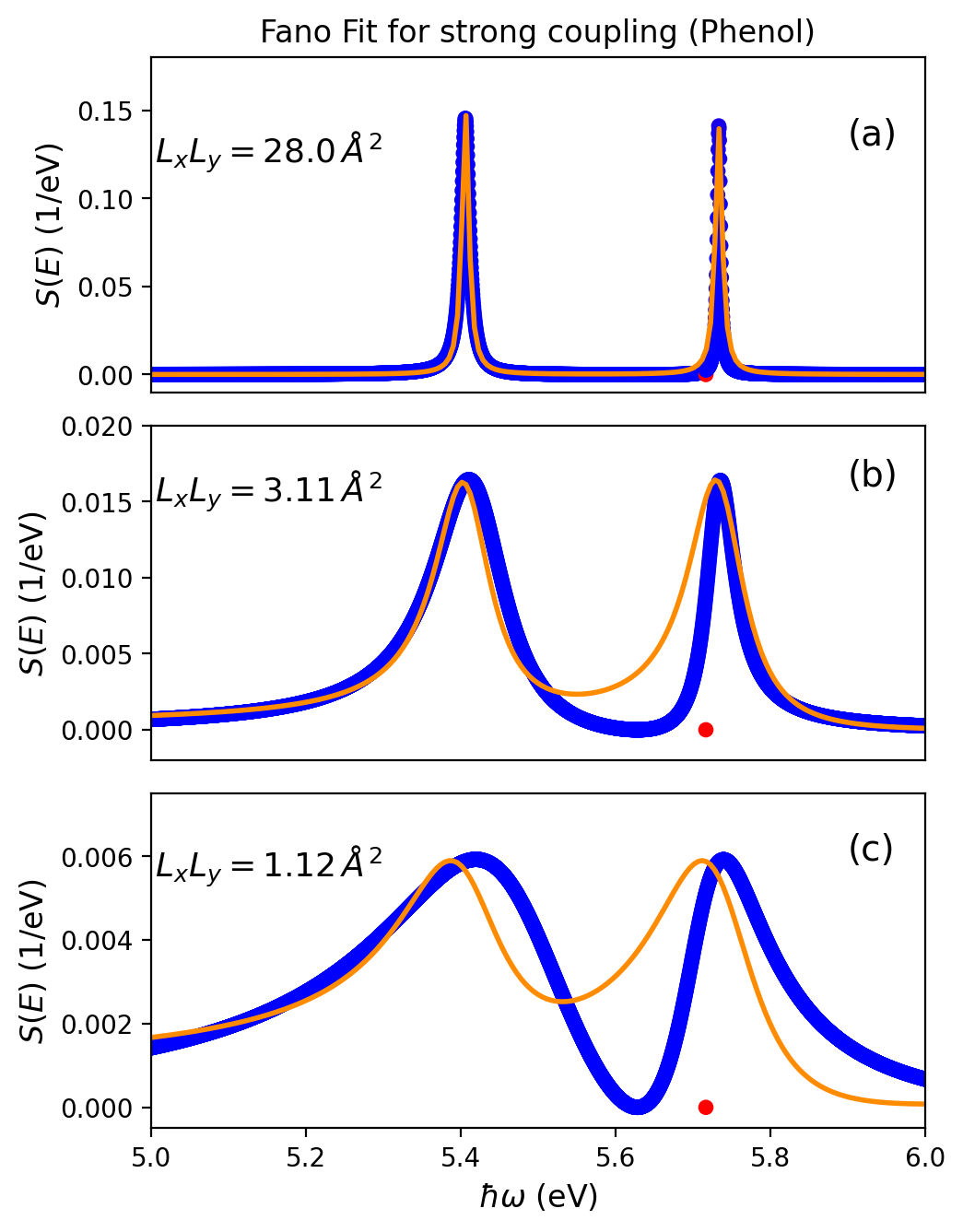}}
\caption{Calculated and Fano fit (orange line) to the polaritonic peaks in Fig.~(\ref{fig:phenol-polaritonic-spectrum}) characterizing a transition from symmetric to asymmetric line shapes. (a) Symmetric Lorentzian line shape occur in the weak coupling to the photon continuum. Panels (b) and (c) show the asymmetric Fano line shapes that emerge due to strong coupling to the photon continuum. The Fano fit become less accurate for stronger coupling.}
\label{fig:phenol-polaritonic-spectrum-fit}
\end{figure}

In the second study, in addition to the modes that sample the electromagnetic vacuum, an additional mode is strongly coupled to an electronic excitation (the peak around $5.6$~eV) that results to the Rabi splitting into the lower and upper polariton peaks. Including an extra mode that couple strongly can be realized, for example, in crossed cavity setups~\cite{heinz2021,brekenfeld2020}. We show the results of this calculation in Fig.~(\ref{fig:phenol-polaritonic-spectrum}). Typically, in this strong coupling regime, the continuum of weakly interacting modes constitute the line width of the excitations and also represent dissipation channels in the coupled system~\cite{welakuh2021,flick2019}. Interestingly, we find in Fig.~(\ref{fig:phenol-polaritonic-spectrum} a) a larger broadening of the lower polaritonic peak when compared to that of the upper polariton (see Fig.~(\ref{fig:phenol-polaritonic-spectrum-fit} a) for details) which implies the lower polariton state has a longer radiative lifetime. When we effectively enhance the light-matter coupling strength, this results to a transition of the initially Lorentzian profiles for the polaritonic excitations into Fano resonances as shown in Fig.~(\ref{fig:phenol-polaritonic-spectrum-fit} b,c). 
\begin{table}[bth]
\begin{tabular}{ | c | c | c | c | }
	\hline
	$\; L_{x}L_{y}$ (\AA$^{2}$) & $\; \varphi$ (rad) &  $\; \hbar\Gamma$ (eV)  & $ \; q$  \\
	\hline \hline
	28.0  &    $\pi$     &   0.019   &  -8.17$\times 10^{15}$      \\
	\hline
	3.11  &  $0.972\pi$  &   0.088   &   -11.43    \\
	\hline
	1.12  &  $0.917\pi$  &   0.161   &   -3.73     \\
	\hline
\end{tabular}
\caption{Fitted values for the polaritonic case of the phase shift $\varphi$, line width $\hbar\Gamma$ and Fano parameter $q$ for the different cases of changing the coupling strength to the continuum by varying the cavity mode volume via $L_{x}L_{y}$. This corresponds to the Fano fit in Fig.~\ref{fig:phenol-polaritonic-spectrum-fit}.}
\label{tab:polaritonic-fit}
\end{table}
To prove that the asymmetric line shapes in Fig.~(\ref{fig:phenol-polaritonic-spectrum-fit}) are indeed Fano resonances, we again apply the Fano formula of Eq.~(\ref{eq:cross-section-Fano-formula}) to the polaritonic peaks between the energy range $5$ to $6$~eV (the fitted values are given in Tab.~(\ref{tab:polaritonic-fit})).
In Fig.~(\ref{fig:phenol-polaritonic-spectrum-fit} a), the Fano fit is in good agreement with the calculated spectrum as it characterizes the Lorentzian line shapes corresponding to the lower and upper polariton peaks. The symmetric line shapes are equally due to the negligible transition amplitudes for coupling weakly to the continuum as in electronic case. In contrast to Fig.~(\ref{fig:phenol-polaritonic-spectrum-fit} a), the Fano fit deviates from the calculated spectra for increasing coupling to the continuum as shown in Fig.~(\ref{fig:phenol-polaritonic-spectrum-fit} b,c). This deviation indicates the need for a self-consistent treatment of light-matter interaction which cannot be captured by perturbative approaches. We find that the line width of the polaritonic peaks are broader and leads to a Purcell enhancement larger than the electronic case which could be useful in an optical quantum memory devices~\cite{lvovsky2009}. Equally, even for strong coupling we expect to find a Lamb shift of polaritonic excitations due to interactions with vacuum fluctuations. It is likely the case that the Lamb shift will increase when the vacuum density of states gets perturbed as in this case as seen in Fig.~(\ref{fig:phenol-polaritonic-spectrum-fit} c). The reverse process of switching from a Fano to a Lorentzian profile can be achieved by reducing the coupling strength (which implies increasing the cavity volume) to the continuum of modes. Such a change reverts the modified photonic density of states to that of the free-space emission. A key interference effect that arises in setups that are associated with Fano resonances is electromagnetically induced transparency (EIT). This effect is a quantum interference phenomenon that allow light pulses through an otherwise opaque medium~\cite{harris1997}. EIT is a special case of Fano resonance that occur due to a destructive interference between the discrete resonant state and a continuum state (i.e., when frequencies match) such that $q=0$, which describes the resonant suppression of absorption~\cite{limonov2017}. Our description of the coupled system captures the EIT phenomenon as discussed in detail in App.~\ref{app:EIT-section}. The different examples presented in this work make the cavity setup a versatile approach for realizing distinct spectroscopic signatures of symmetric Lorentz and asymmetric Fano resonances within the framework of non-relativistic quantum electrodynamics.

\section{Summary and outlook}
\label{sec:summary-outlook}

In conclusion, we present an approach that allows switching between two spectroscopic signatures associated with symmetric Lorentzian and asymmetric Fano line shapes. To achieve such control, we consider a setting for realizing strong light-matter interaction where the photonic environment in which the matter subsystem is immersed is of a multimode nature. In this cavity setup, we show that by strongly coupling the bare matter and hybridized system to the continuum of transverse electromagnetic modes, the initial spectroscopic signatures of symmetric Lorentzian line shapes are tuned into asymmetric Fano line shapes. We further confirm that these asymmetric line shapes are indeed Fano resonances by fitting the Fano formalism to the calculated spectra where we find good agreement for the case of discrete electronic states interacting strongly with the continuum. In the case where we coupled polaritonic states to the continuum of modes, we find important differences between the fit and \emph{ab initio} results, highlighting the need for a first-principles description of the interference phenomenon of Fano resonances where the perturbative Fano formalism breaks down. From the Fano fit, we are able to characterize the degree of asymmetry of the Fano resonances as well as deduce directly the finite radiative lifetimes of excited states including the Purcell enhancement of spontaneous emission due to a change in the photonic density of states. We also showed that our cavity setup for realizing interference effects also captures the special case of Fano resonance of EIT. This work presents a step forward in the physics of Fano resonances as the asymmetric signatures can be tuned and controlled by effectively enhancing the light-matter coupling to the electromagnetic continuum. Finally, the results presented here show the new perspectives that \emph{ab initio} QED approaches now provide to analyze and predict quantum nanophotonics in regimes of strong coupling.

\section{Acknowledgments}
We acknowledge helpful discussions with Marissa Weichman and Stephane Kena-Cohen. This work is supported by the Office of Naval Research (ONR) MURI Program under grant number ONR N00014-21-1-2537 as well as the Quantum Science Center (QSC), a National Quantum Information Science Research Center of the U.S. Department of Energy (DOE). P.N. acknowledges support as a Moore Inventor Fellow through Grant No. GBMF8048 and gratefully acknowledges support from the Gordon and Betty Moore Foundation as well as support from a NSF CAREER Award under Grant No. NSF-ECCS-1944085.

\appendix

\section{Electromagnetically induced transparency (EIT)}%
\label{app:EIT-section}

\begin{figure}[bth]
\centerline{\includegraphics[width=0.5\textwidth]{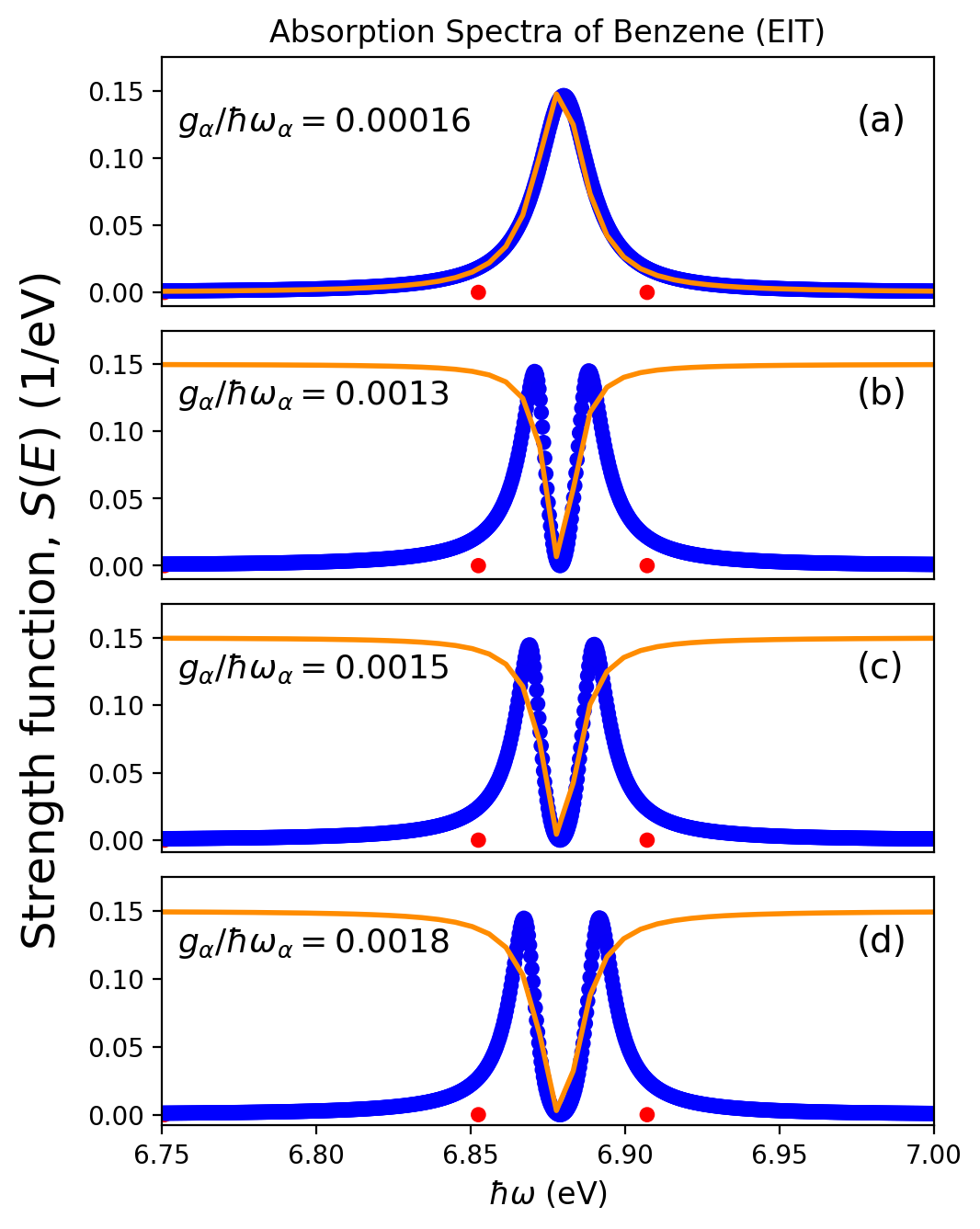}}
\caption{(a) The absorption spectra of a single benzene molecule capturing the $\pi-\pi^{*}$ excitation. Panels (b)-(d) show the onset of a window of transparency for an increasing coupling strength to a continuum mode. The quasi-Lorentzian antiresonance Fano fit characterizes the transmission spectra at the transparency window.}
\label{fig:benzene-EIT}
\end{figure}

Electromagnetically induced transparency (EIT) is a quantum interference effect that allows the propagation of light pulses through an otherwise opaque medium~\cite{harris1990,harris1997}. EIT is a special case of Fano resonance that occur due to a destructive interference between the discrete resonant state and a continuum state (i.e., when frequencies match) such that the Fano parameter vanishes at $q=0$, which describes the resonant suppression of absorption~\cite{limonov2017}. One major application of EIT is in light delay and storage, where light propagating through a medium is slowed down~\cite{haugland2020} and even stopped~\cite{liu2005,welakuh2017}. We now consider how to realize EIT in the cavity setup depicted in Fig.~(\ref{fig:phenol-in-cavity}). In the following, we replace the phenol molecule with a single benzene molecule. By so doing, we highlight that the interference effects presented in this work is independent of the matter system. To realize EIT in the current setup, we enhance the coupling of one of the sampled continuum of modes to the $\pi-\pi^{*}$ excitation of benzene around $6.855$~eV. The rest of the continuum modes couple weakly as discussed in Sec.~\ref{sec:general-framework} where the cavity area is $L_{x}L_{y}=28.0$ \AA$^{2}$. The results of this calculation are shown in Fig.~(\ref{fig:benzene-EIT}) where we capture the $\pi-\pi^{*}$ excitation of benzene around $6.855$~eV as shown in Fig.~(\ref{fig:benzene-EIT} a). When the frequency of the enhanced cavity mode matches that of the $\pi-\pi^{*}$ excitation, this induces a window of transparency at which the system can no longer absorb light as shown in Fig.~(\ref{fig:benzene-EIT} b-d).
The Fano fit is in good agreement to the computed spectra and the fitted values are given in Tab.~(\ref{tab:EIT-fit}).
\begin{table}[bth]
\begin{tabular}{ | c | c | c | c | c | }
	\hline
	$g_{\alpha}$ (eV) & $g_{\alpha}/\hbar\omega_{\alpha}$ & $\; \varphi$ (rad) &  $\; \hbar\Gamma$ (eV)  & $ \; q$  \\
	\hline \hline
	0.0011  &   0.00016 &    $\pi$  &   0.019   &  -8.16$\times 10^{15}$      \\
	\hline
	0.0087  &   0.0013  &  $\pi/2$  &   0.011   &   0    \\
	\hline
	0.010   &   0.0015  &  $\pi/2$  &   0.014   &   0     \\
	\hline
	0.0122  &   0.0018  &  $\pi/2$  &   0.016   &   0    \\
	\hline
\end{tabular}
\caption{Fitted values for EIT for the phase shift $\varphi$, line width $\hbar\Gamma$ and Fano parameter $q$ for the different coupling strength of an enhanced resonant bath mode. These values corresponds to the Fano fit in Fig.~(\ref{fig:benzene-EIT}).}
\label{tab:EIT-fit}
\end{table}
Upon increasing the coupling strength leads to an increase in the transparency window. With these results, EIT phenomenon is now for the first time realized from first-principles using QEDFT linear-response framework and our results show the new perspectives that can be achieved with \emph{\emph{\emph{\emph{ab initio}}}} QEDFT methods~\cite{flick2019,flick2018,welakuh2022}. We note that EIT has also been studied using a time-dependent density-functional theory (TDDFT) approach that includes a local potential which accounts for the coupling to the electromagnetic continuum~\cite{schaefer2021a}. 

\begin{figure}[bth]
\centerline{\includegraphics[width=0.5\textwidth]{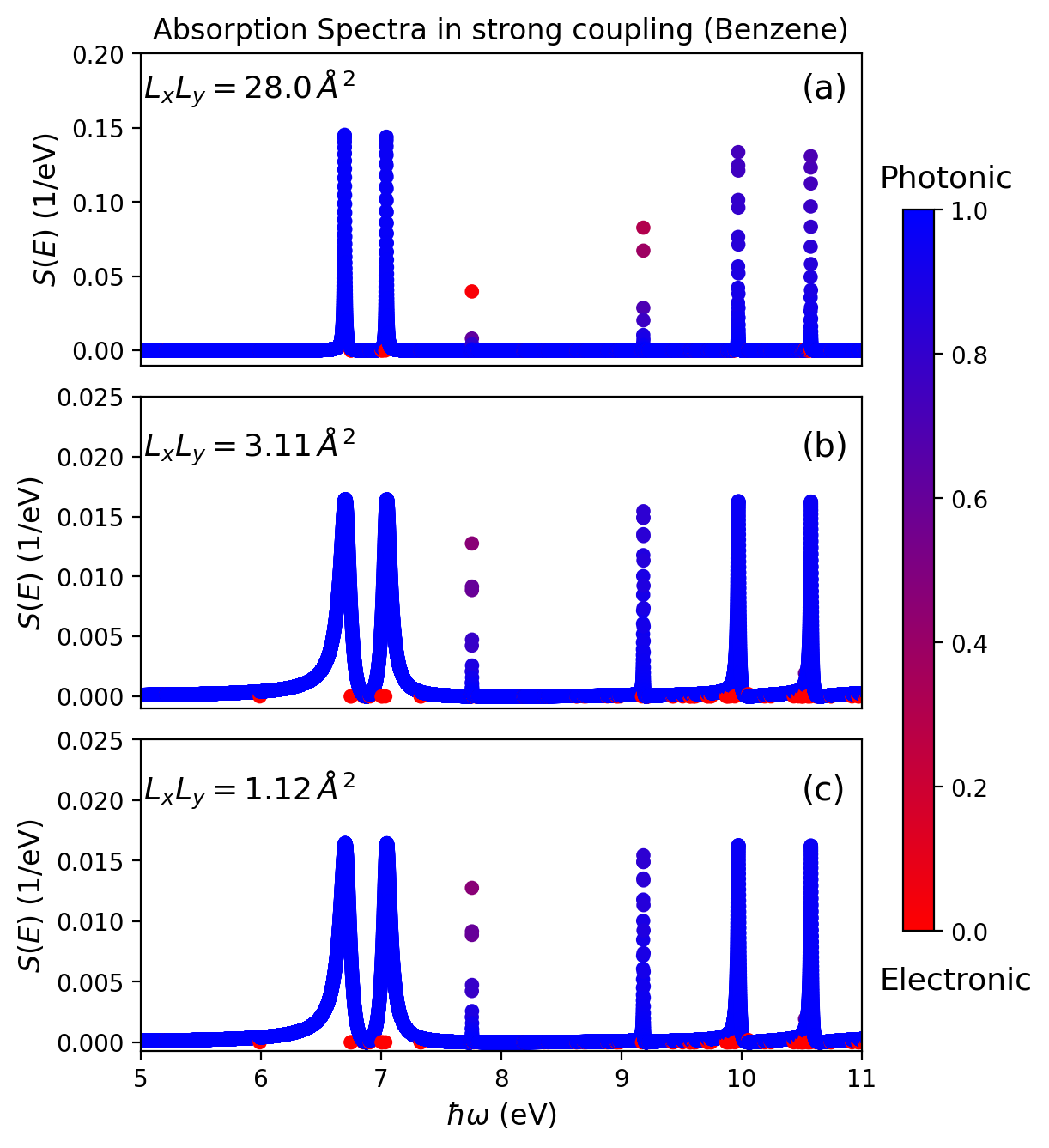}}
\caption{(a) \emph{Ab initio} calculation of the absorption spectrum of benzene capturing the lower and upper polaritonic peaks within the range $6.5$ to $7.5$~eV due to strong coupling to the $\pi-\pi^{*}$ excitation. The line shapes are symmetric Lorentzians. Panels (b) and (c) shows how both polaritonic and electronic excitations can be tuned into Fano resonances by strongly coupling to the photon continuum.}
\label{fig:benzene-polaritonic}
\end{figure}

It is important to note that the splitting of the absorption peak that occur at the transparency window (see Fig.~(\ref{fig:benzene-EIT} b-d)) differs from the strong coupling regime that leads to the formation of polariton peaks (see Fig.~(\ref{fig:benzene-polaritonic})). The difference between these two effects is that, EIT occurs when the decay rate from the resonant state is greater than the coupling strength $g_{\alpha}$ (see Sec.~\ref{sec:numerics} for details) and $g_{\alpha}$ is greater than the decay rate of the continuum while for strong coupling, $g_{\alpha}$ is greater than both decay rates~\cite{limonov2017}. To demonstrate this, we compare the values of the first column to the fourth column of Tab.~(\ref{tab:EIT-fit}) where we find that the EIT condition is satisfied since $g_{\alpha}$ will be greater than the decay rate of the continuum. The strong coupling case is immediately satisfied since $g_{\alpha}=0.174$~eV is greater than the respective decay rates of the EIT case. Therefore, increasing the coupling to the resonant mode allows a transition between EIT to the strong coupling regime.

\section{Numerical details}
\label{sec:numerics}

We now give the numerical details of the calculations presented in this work. To calculate the electronic structure of the phenol molecule, we use a cylindrical real space grid of $8$~{\AA} length with the radius of $6$~{\AA} in the $x$-$y$ plane, and a spacing of $\Delta x = \Delta y = \Delta z = 0.22$~\AA. We explicitly describe the 36 valence electrons which amounts to 18 doubly occupied orbitals, while the core atoms are considered implicitly by the local-density approximation (LDA) Troullier-Martins pseudopotentials~\cite{troullier1993}. In the excited state manifold, we include $500$ unoccupied states in the pseudo-eigenvalue calculation. This number amounts to $N_{v}*N_{c}=18*500=9000$ pairs of occupied-unoccupied states where $N_{v}$ and $N_{c}$ denote the number of occupied and unoccupied Kohn-Sham orbitals, respectively. Using the electron-photon Casida equation, the resulting dimension of the coupled but truncated matrix is $\left((N_{v}*N_{c}+M)\times(N_{v}*N_{c}+M)\right)$ where we sampled $M=100,000$ photon modes in the first example and $M=100,001$ in the second case. To describe the electron-electron and electron-photon interaction in the linear-response QEDFT framework~\cite{flick2019}, we apply the adiabatic LDA (ALDA) to the Hartree exchange-correlation kernel (i.e. $f^{n}_{\text{Hxc}} \rightarrow f^{n}_\text{Hxc,ALDA}$) and the photon random-phase approximation to the photon exchange-correlation kernels (i.e. $f^{n}_{\text{pxc}}, f^{q_{\alpha}}_{\text{pxc}} \rightarrow f^{n}_\text{p}, f^{q_{\alpha}}_\text{p}$), respectively. We define the electron-photon coupling from the bilinear interaction term of Eq.~(\ref{eq:el-pt-hamiltonian}) as is Ref.~\cite{welakuh2021t}:
\begin{align}
g_{\alpha}^{(ij)} = \sqrt{\frac{\hbar\omega_{\alpha}}{\epsilon_{0}V}} \langle \varphi_{i}|\textbf{e}_{\alpha}\cdot\hat{\textbf{R}}|\varphi_{j} \rangle \, ,
\end{align}
where $|\varphi_{j} \rangle$ are the eigenstates of the bare matter system and $\langle \varphi_{i}|\textbf{e}_{\alpha}\cdot\hat{\textbf{R}}|\varphi_{j} \rangle$ is the transition dipole moment. We compute the $x$-contribution of the dipole strength function~\cite{flick2019} with respect to the polarization of the cavity modes.

We numerically describe the Benzene molecule coupled to photons in App.~\ref{app:EIT-section} using the same details as the phenol molecule presented here. The only difference is that benzene has 30 valence electrons which amounts to 15 doubly occupied orbitals. In this case, the number of occupied-unoccupied pairs is $N_{v}*N_{c}=15*500=7500$.

\vspace{10em}

\bibliography{01_light_matter_coupling} 

\end{document}